\documentclass[prl,aps,twocolumn,showpacs]{revtex4-1} 
\usepackage{amsmath,amssymb,amsthm}
\usepackage{graphicx}
\usepackage{subfigure}
\usepackage{amsmath}
\usepackage{amsfonts,amssymb}
\usepackage{bm}
\usepackage{float}
\usepackage{color}
\usepackage{sidecap}
\usepackage{dsfont}
\allowdisplaybreaks
\usepackage{soul}
\setstcolor{red}
\usepackage[normalem]{ulem}
\usepackage{hyperref}
\hypersetup{colorlinks=true,breaklinks,urlcolor=blue,linkcolor=blue,citecolor=blue}

\begin{document}

\title{Higher-order topological insulators and semimetals in generalized Aubry-Andr\'e-Harper models}
\author{Qi-Bo Zeng$^{1}$}
\thanks{These authors contributed equally to this work.}
\author{Yan-Bin Yang$^{1}$}
\thanks{These authors contributed equally to this work.}
\author{Yong Xu$^{1,2}$}
\email{yongxuphy@tsinghua.edu.cn}
\affiliation{$^{1}$Center for Quantum Information, IIIS, Tsinghua University, Beijing 100084, People's Republic of China}
\affiliation{$^{2}$Shanghai Qi Zhi Institute, Shanghai 200030, People's Republic of China}

\begin{abstract}
Higher-order topological phases of matter have been extensively studied in various areas of physics. While the Aubry-Andr\'e-Harper model provides a paradigmatic
example to study topological phases, it has not been explored whether a generalized Aubry-Andr\'e-Harper model can
exhibit a higher-order topological phenomenon. Here, we construct a two-dimensional higher-order
topological insulator with chiral symmetry based on the Aubry-Andr\'e-Harper model. We find the coexistence
of zero-energy and nonzero energy corner-localized modes.
The former is protected by the quantized quadrupole moment, while the latter by
the first Chern number of the Wannier band. The nonzero-energy mode can also be viewed as the consequence of
a Chern insulator localized on a surface. More interestingly, the non-zero energy corner mode can lie in the continuum of extended bulk states and form a bound state in the continuum of higher-order topological systems. We finally propose an experimental scheme to realize our model
in electric circuits. Our study opens a door to further study higher-order
topological phases based on the Aubry-Andr\'e-Harper model.
\end{abstract}

\maketitle
\date{today}
Recently, topological phases of matter have made substantial progress
due to the discovery of higher-order topological insulators (HOTIs).
The HOTIs are a generalization of traditional first-order topological insulators
so that for an $n$-dimensional system, $(n-m)$-dimensional edge
states with $m\ge 2$ emerge. To date, there have been extensive studies of HOTIs from various fields
of physics~\cite{Fritz2012PRL,ZhangFan2013PRL,Taylor2017Science,Slager2015PRB,Brouwer2017PRL,FangChen2017PRL,PengYang2017PRB,Taylor2017PRB,FangChen2017arXiv,Wan2017arXiv,
Bernevig2018SA,Ezawa2018PRL,WangZhong2018PRL,ZhangFan2018PRL,Ryu2018PRB,Brouwer2018PRB,Ortix2018PRB,Taylor2018PRB,Fulga2018PRB,Watanabe2018PRB,You2018PRB,
Taylor2018PRB(R),Nori2018PRB,Vincent2018arXiv,Brouwer2019PRX,Wakabayashi2019PRL,Liu2019PRL,DasSarma2019PRL,Yan2019PRL,Zhijun2019PRL,Lee2019PRL,Roy2019PRB,Dubinkin2019PRB,Chen2019arXiv,Pixley2019arXiv,Khalaf2019arXiv,
DasSarma2019arXiv,Wieder2019arXiv,Xu2019arxiv} and several predicted topological phases have been experimentally observed~\cite{Huber2018Nature,Bahl2018Nature,Imhof2018Nat,Jianhua2019NP,Neupert2018NP,Baile2019NatMat,Khanikaev2019NatMat,
Hafezi2019Natphoton,Baile2019PRL,Xiangdong2019arXiv,Khanikaev2019arXiv}. Two typical HOTIs include two-dimensional (2D) quadrupole topological insulators
with zero-energy corner-localized states characterized by a quantized quadrupole moment~\cite{Taylor2017Science,Taylor2017PRB} and three-dimensional (3D) second-order topological insulators with
nonzero-energy chiral hinge states characterized by the Chern number of Wannier bands~\cite{Bernevig2018SA}. However, whether a system can host both zero-energy corner modes and nonzero-energy chiral modes is still elusive.

The Aubry-Andr\'e-Harper (AAH) model~\cite{Aubry1980,Harper1955}, a one-dimensional (1D) system with on-site cosinusoidal modulations,
plays a crucial role in studying the Anderson localization and quasicrystals~\cite{Siggia1983PRL,Kohmoto1983PRL,DasSarma1988PRL,DasSarma1990PRB,DasSarma2009PRA,DasSarma2010PRL}. Recently, it has been found that such a model can also exhibit topological properties~\cite{Lang2012PRL,Zilberberg2012PRL,Zilberberg2012PRLb,Ganeshan2013PRL,Cai2013PRL,Chong2015PRB,Hu2016PRB,Zeng2016PRB,
Yi2018PRA,Zeng2020PRB}. For instance, the model can be mapped to a 2D quantum Hall insulator that hosts nonzero energy topological edge modes characterized by the first Chern number~\cite{Lang2012PRL,Zilberberg2012PRL,Zilberberg2012PRLb}. This model has been further generalized to an off-diagonal AAH model
with modulations in the hopping terms~\cite{Ganeshan2013PRL,Zeng2020PRB},
supporting topologically protected zero-energy modes. Such a model can support the coexistence of topological
zero-energy and nonzero energy modes~\cite{Ganeshan2013PRL}.
It is natural to ask whether we can construct a higher-order topological system based on the 1D AAH model so that the coexistence of these two topological phenomena can be observed.

Another interesting phenomenon concerns the existence of localized states in the continuous bulk spectrum of a system, which are known as bound states in the continuum (BICs). BICs have been widely investigated
in a wide range of physical systems, such as optical systems~\cite{Longhi2008PRA,Longhi2009PRA,Dreisow2009OL,Plotnik2011PRL},
plasmonic-photonic systems~\cite{Paddon2000PRB,Pacradouni2000PRB,Ochiai2001PRB,Fan2002PRB,Hsu2013Nat,Zhen2014PRL}, 
acoustics~\cite{Parker1966JSV,Parker1967JSV,Cumpsty1971JSV,Koch1983JSV,Parker1989,Evans1994}, 
quantum dots~\cite{Guevara2003PRB,Orellana2004PRB,Guevara2006PRB,Voo2006PRB,Gong2009JPCM} and 
water waves~\cite{Ursell1951,Jones1953,Callan1991,Retzler2001,Cobelli2009EPL,Cobelli2011}.
Recently, it has been shown that BICs can exist in topological systems~\cite{Yang2013NatCom,Xiao2017PRL,ZGChen2019PRB,Benalcazar2019arxiv}.
Similar to other classical systems, an electric network system has recently been proven to be a powerful platform
to simulate topological phenomena~\cite{Lee2018CommPhys,Lu2019PRB,Yang2019PRL,Zeng2020PRB,Imhof2018Nat}. However, whether such systems can support BICs has not yet been explored. We will report a proposal with an electric circuit to realize a higher-order AAH model supporting
BICs.

In this Rapid Communication, we address these important questions by constructing a 2D AAH lattice model based on 1D commensurate off-diagonal AAH models. We find that (i) zero-energy and nonzero-energy corner-localized states can coexist in the system.
The former is protected by a quantized quadrupole moment and the latter is protected by the Chern number
of the boundary bands as well as the Chern number of the Wannier bands.
More interestingly, these nonzero-energy corner modes can exist in a continuous bulk spectrum forming bound states in the continuum.
(ii) After transforming the system into a 3D lattice model,
the system becomes a topological semimetal at half filling with gapless points located either on the surfaces or in the bulk.
(iii) For the 3D system, besides the topological zero-energy modes localized at the hinges, the system also exhibits nonzero-energy chiral hinge modes arising from
the existence of Chern bands on the surfaces.
(iv) We finally propose an experimental scheme to simulate the 2D AAH model using electric circuits
and demonstrate the existence of BICs in the form of voltage waves in the circuit.

\begin{figure*}[t]
  \includegraphics[width=7.0in]{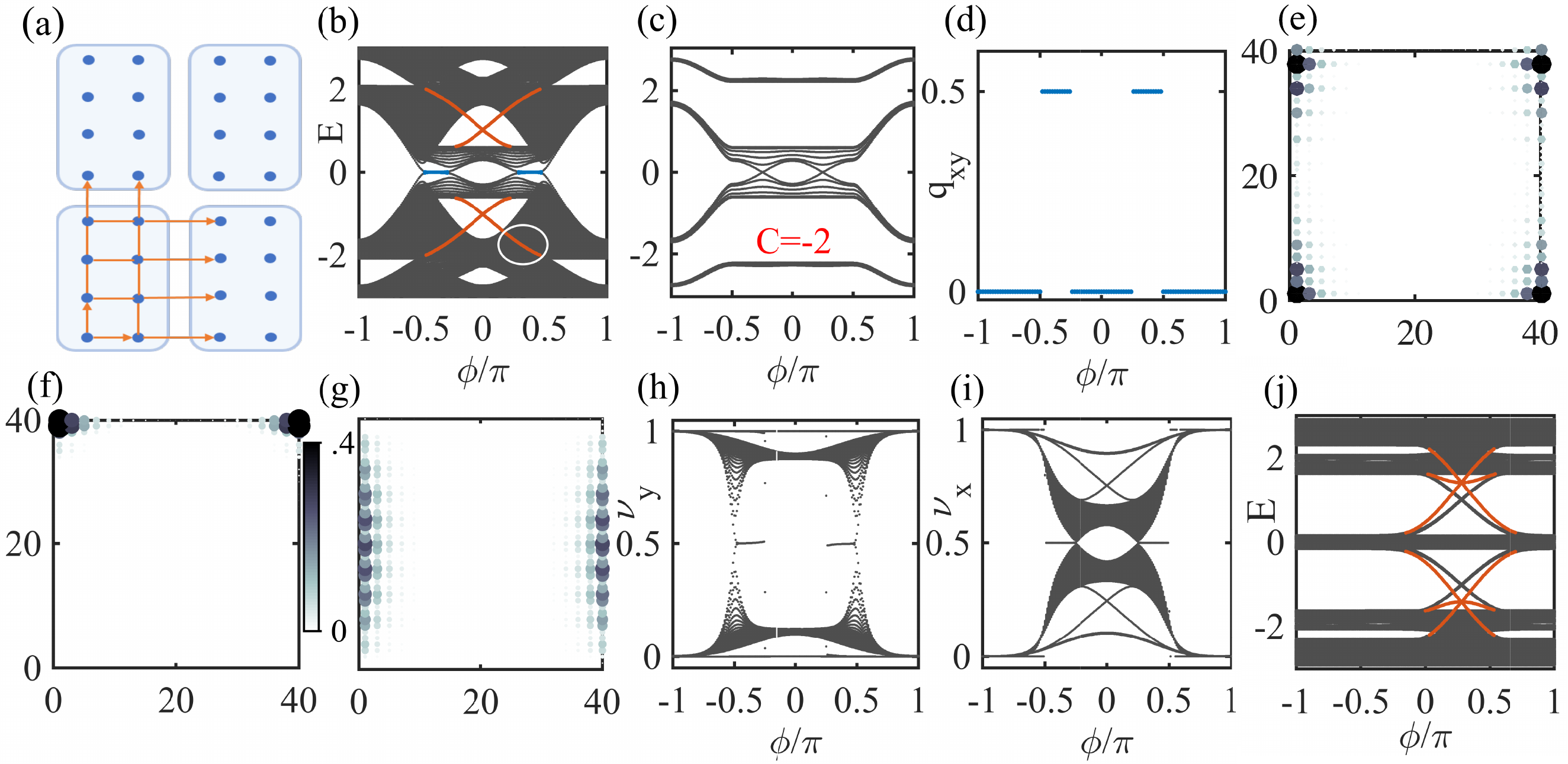}
  \caption{(Color online) (a) Schematics of the tunneling for the 2D AAH lattice model with $\alpha_x=1/2$ and $\alpha_y=1/4$. (b) The energy spectrum obtained under open boundary conditions along both $x$ and $y$ directions. The lattice size is $40 \times 40$. The blue and red lines describe the zero-energy and nonzero energy topological corner modes, respectively. The nonzero-energy corner modes highlighted by the white circle exist in a continuous bulk band. (c) The boundary bands obtained under periodic (open) boundary conditions along $y$ ($x$). The red number denotes the Chern number contributed by the boundary bands. (d) The quadrupole moment in units of $e$ as a function of $\phi$ at half filling. (e)-(g) display the 
  spatial distribution of the wave-function amplitude for a zero-energy corner mode, the nonzero-energy corner modes, and the states for the boundary bands at $\phi=0.35\pi$, respectively. The Wannier spectra $\nu_y$ (h) and $\nu_x$ (i) are computed under open boundary conditions along $x$ and $y$ directions, respectively. The presence of edge states around $\nu_x=0.5$ and $\nu_y=0.5$ in the Wannier spectra shows the existence of quantized edge polarizations.
  The Wannier band $\nu_x$ (similarly for $\nu_y$) refers to the eigenvalues $2\pi\nu_x$ of the Wannier Hamiltonian
  $H_{\mathcal{W}_x}$ defined as $\mathcal{W}_x\equiv e^{iH_{\mathcal{W}_x}}$, where
  $\mathcal{W}_x=F_{x,k_x+(N_x-1)\delta k_x}\cdots F_{x,k_x}$ is the Wilson loop with
  $[F_{x,k_x}]^{mn}=\langle u_{k_x+\delta k_x}^m|u_{k_x}^n\rangle$~~\cite{Taylor2017Science,Taylor2017PRB}
  and $|u_{k_x}^n\rangle$ is the $n$th occupied eigenstate
  of our system and $\delta k_x=2\pi/L_x$, with $L_x$ being the number of unit cells along $x$. Here, $t=1$, $\lambda_x=\lambda_y=0.8$, $\alpha_x=1/2$, and $\alpha_y=1/4$. (j) shows the energy spectrum for the incommensurate case with $\alpha=(\sqrt{5}-1)/2$, where only nonzero-energy topological corner modes exist.}
\label{fig1}
\end{figure*}

\emph{Model Hamiltonian}.--- We start by considering the following 2D lattice model [see Fig.~\ref{fig1}(a) for a schematic],
\begin{eqnarray}\label{H}
  H=\sum_{i,j} [t_{(i,j),(i+1,j)} && \hat{c}_{i,j}^\dagger \hat{c}_{i+1,j} + \nonumber \\
  &&t_{(i,j),(i,j+1)} \hat{c}_{i,j+1}^\dagger \hat{c}_{i,j} + H.c.],
\end{eqnarray}
where $c_{i,j}^\dagger$ ($c_{i,j}$) represents the creation (annihilation) operator for a spinless particle at site $(i,j)$. The hopping amplitudes along the $x$ and $y$ directions are respectively modulated as
\begin{align*}\label{}
  t_{(i,j),(i+1,j)} &= (-1)^j t_x [1+\lambda_x \cos (2\pi \alpha_x i + \phi_x)],\\
  t_{(i,j),(i,j+1)} &= t_y[1+\lambda_y \cos (2\pi \alpha_y j + \phi_y)],
\end{align*}
where $\alpha_x =p_x/q_x$, $\alpha_y=p_y/q_y$ with $p_x$ and $q_x$ ($p_y$ and $q_y$) being mutually prime positive integers,
$t_x$ ($t_y$) is the hopping strength between the nearest-neighboring lattice sites along the $x$ ($y$) direction, and $\lambda_x$ and $\lambda_y$ are the amplitudes of the modulations whose periods are determined by $\alpha_x$ and $\alpha_y$. Without loss of generality, we set $t_x=t_y=t$ and $\phi_x=\phi_y=\phi$. The 2D lattice is actually composed by 1D commensurate off-diagonal AAH models. If $\alpha_x=\alpha_y=1/2$, the system exhibits zero-energy corner modes with a quantized electric quadrupole
moment, similar to the model in~\cite{Taylor2017Science}.

With periodic boundary conditions in both directions, we can write the Hamiltonian in Eq.(\ref{H}) in the momentum space as~\cite{Okugawa2019arxiv}
\begin{equation}\label{Hk}
  H(k_x,k_y)=H_x(k_x) \otimes \Pi_y + I_x \otimes H_y(k_y),
\end{equation}
where $H_i(k_i)$ ($i=x,y$) is the Hamiltonian of the 1D AAH model along the $i$ direction with $k_i \in [0,2\pi)$. Such
a 1D Hamiltonian is a $q_i\times q_i$ matrix, which, when $q_i$ is even, has chiral symmetry represented by a $q_i\times q_i$ diagonal matrix $\Pi_i=\textrm{diag}(\begin{array}{ccccc}
              1 & -1 & \cdots & 1 & -1
            \end{array})$,
i.e., $\Pi_i H_i \Pi_i^\dagger=-H_i$. $I_x$ is an identity matrix of size $q_x$. Clearly, the 2D Hamiltonian has chiral symmetry, $\Pi H \Pi^\dagger=-H$ with $\Pi=\Pi_x\otimes \Pi_y$.
For the 1D AAH model with $q_i>2$, we can obtain chiral edge modes by tuning the parameter $\phi$, which leads to topological Chern bands. In the following, we will explore whether the 2D lattice composed of such 1D AAH models can give rise to new topological features.

\emph{Case $\alpha_x=1/2$ and $\alpha_y=1/4$}.--- We first consider the model with $\alpha_x=1/2$ and $\alpha_y=1/4$.
Figure~\ref{fig1}(b) shows the energy spectrum of the lattice model under open boundary conditions along both directions. For $\phi \in (-\frac{\pi}{2}, -\frac{\pi}{4})$ and $(\frac{\pi}{4}, \frac{\pi}{2})$, we can observe zero-energy modes in the gap marked by the blue lines. They are four-fold degenerate and localized at the four corners of the 2D lattice, as illustrated in Fig.~\ref{fig1}(e).
The topological zero-energy corner states signify the existence of quadrupole moments in the system. To prove this,
we have numerically calculated the quadrupole moment of the model as a function of $\phi$ based on the following formula~\cite{Wheeler2018arxiv, Kang2018arxiv}
\begin{equation}\label{quadurpole}
  q_{xy}=\frac{e}{2\pi}\text{Im}\log \text{det} U,
\end{equation}
where
$U_{nm}=\langle u_n|e^{i2\pi xy/(L_xL_y)}|u_m\rangle$
with $|u_m\rangle$ being the $m$th occupied eigenstate and $L_x$ ($L_y$)
being the length of our system in the $x$ ($y$) direction relative
to the unit cells. We perform our calculation under periodic boundary conditions.
Due to the chiral symmetry, the quadrupole moments are quantized to 0 or
$e/2$ modulo one.

In Fig.~\ref{fig1}(d), we present the quadrupole moment with respect to $\phi$, showing that
the quadrupole moments are equal to $e/2$ in the parameter regimes where the zero-energy
corner states emerge. This implies that the zero-energy states are characterized by
the quantized quadrupole moment. We can also understand the
presence of zero-energy corner states from the topology of the 1D AAH model~\cite{Okugawa2019arxiv}.
Consider a system with a cylinder geometry with open boundaries along $x$ and periodic boundaries along $y$.
If $H_x$ is topologically nontrivial, there are two zero-energy edge states for $H_x$ denoted as
$|u_{xL}\rangle$ and $|u_{xR}\rangle$ localized at the left and right edge, respectively.
Clearly, the effective Hamiltonian of our 2D system at one edge around zero energy
is $H_y(k_y)$, resulting in two zero-energy corner states if $H_y$ is topologically nontrivial and
we impose open boundary conditions along $y$. Specifically, a state localized at the
top left corner can be written as $|u_{xL}\rangle\otimes |u_{yT}\rangle$, where
$|u_{yT}\rangle$ is the zero-energy edge state of $H_y$ localized at the top edge.
Since $H_x$ and $H_y$ respect the chiral symmetry, their zero-energy edge states can
be characterized by the winding number $W_x$ and $W_y$, respectively.
As a result, their product $W=W_x W_y$ gives another diagnosis of the topological property
for the 2D system, apart from the quadrupole moment.
Since the zero-energy state emerges for $H_x(k_x)$ with $\alpha_x=1/2$ when $\phi \in (-\frac{\pi}{2}, \frac{\pi}{2})$ and
for $H_y(k_y)$ with $\alpha_y=1/4$ when $\phi \in (-\frac{\pi}{2},-\frac{\pi}{4})$ and $(\frac{\pi}{4}, \frac{\pi}{2})$,
the intersection of $\phi$ where both zero-energy states exist corresponds to the parameter region observed in Fig.~\ref{fig1}(d). Due to the chiral symmetry, the zero-energy edge modes of the 1D AAH model are characterized by winding numbers. By combining the two winding numbers along the $x$ and $y$ directions, the zero-energy corner modes can be characterized~\cite{Okugawa2019arxiv}, which is consistent with the quadrupole moment results here.

It is known that the quadrupole moment can change due to the bulk or edge energy gap closure~\cite{Taylor2017Science,Xu2019arxiv}. In our system,
at $\phi=\pm \pi/2$, the change arises from the $y$-normal edge energy gap closure, and at $\phi=\pm \pi/4$,
from the $x$-normal edge energy gap closure. The former (latter) gap closure also leads to the gap closure for the Wannier
band $\nu_y$ ($\nu_x$) [see Figs.~\ref{fig1}(h) and \ref{fig1}(i), respectively].
As a consequence, the edge polarizations along both directions $p_x^{\textrm{edge }}=p_y^{\textrm{edge }}=e/2$ in the topological region with
quadrupole moments. In addition, fractional corner charges appear as $Q^{\text{corner}}=\pm e/2$ at half filling since there are four zero-energy corner modes. This shows that the basic relation
$Q^{\textrm{corner }}=(p_y^{\textrm{edge }}+p_x^{\textrm{edge }}-q_{xy})\text{mod}(1)$ is respected and thus it is a type-I quadrupole topological insulator~\cite{Xu2019arxiv}.

Apart from the zero-energy corner modes, we also observe nonzero-energy topological corner modes in the energy spectrum, as shown by the red lines in Fig.~\ref{fig1}(b). If $\phi$ is viewed as the momentum along $z$, these modes are chiral and we thus refer to 
these modes as chiral modes. 
In the following, we will show that these modes arise from a surface Chern insulator.
By calculating the energy spectrum of the system with periodic and open boundaries along $y$ and $x$, respectively,
we extract the boundary bands [shown in Fig.~\ref{fig1}(c)] based on their localization property~\cite{SM}.
These bands are twofold degenerate with the states localized at either the right or left edge. The nonzero-energy topological modes connecting these boundary bands are characterized by the Chern number defined in the $(k_y, \phi)$ space as
\begin{equation}\label{Chern}
  C_n = \frac{1}{2\pi}\int_0^{2\pi} d\phi \int_{0}^{2\pi}dk_y \Omega_n(k_y,\phi),
\end{equation}
where $\Omega_n=i(\langle \partial_{k_y} \Psi_n |\partial_\phi \Psi_n \rangle -k_y \leftrightarrow \phi)$
with $|\Psi_n \rangle$ being the states in the boundary band extracted from the energy spectrum for a system with
periodic boundaries along $y$. We find $C=-2$ for the lowest bands in Fig.~\ref{fig1}(c) due to the degeneracy, with
each boundary band localized at one of the two edges contributing a Chern number of $-1$.

\begin{figure}[t]
  \includegraphics[width=3.0in]{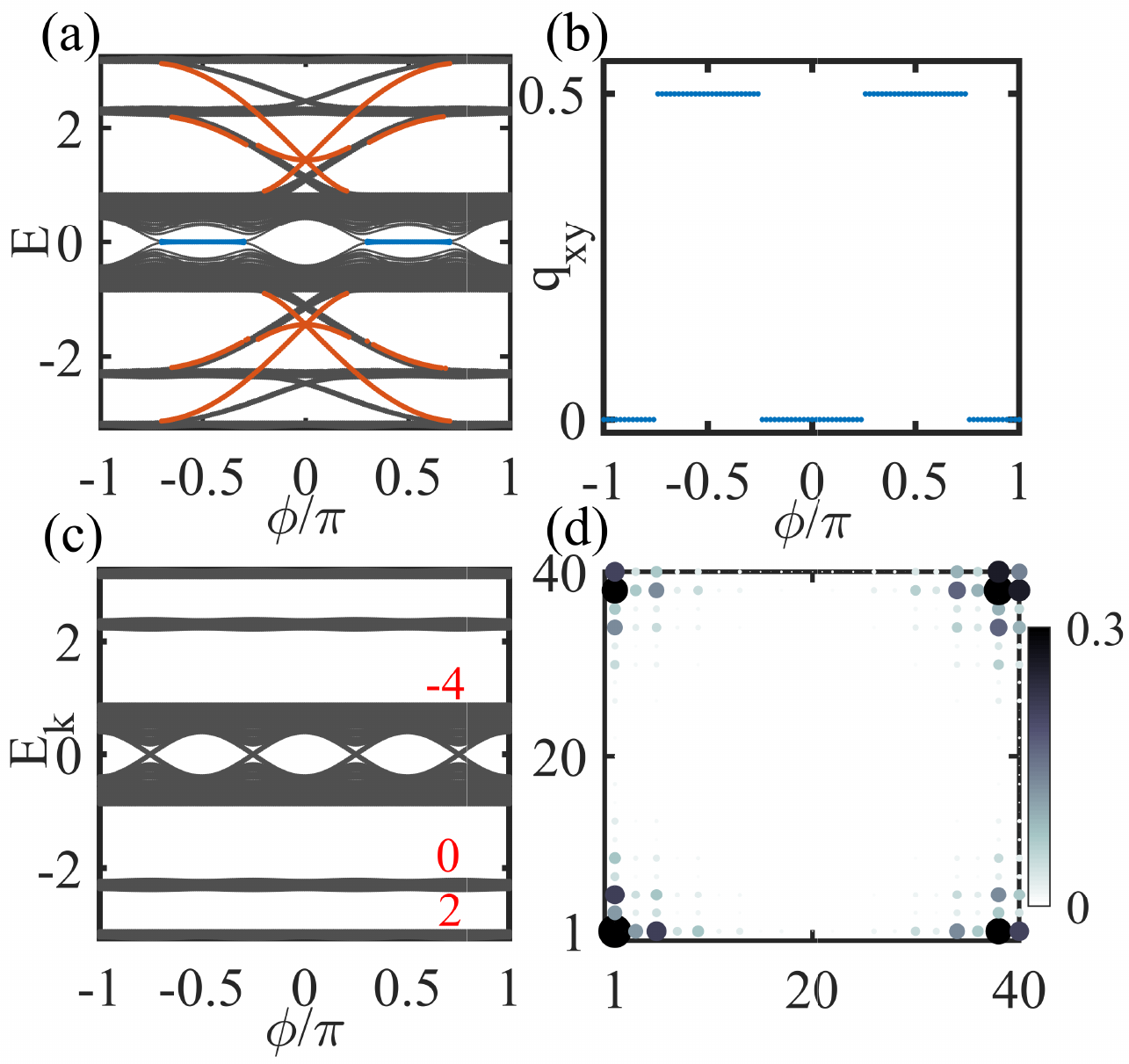}
  \caption{(Color online) (a) The energy spectrum obtained under open boundary conditions along both $x$ and $y$ directions. The lattice size is $40 \times 40$. The red lines describe the chiral corner modes with respect to $\phi$ inside the gap. (b) The quadrupole moments in units of $e$ as a function of $\phi$. (c) The energy spectrum obtained under periodic boundary conditions along both directions. The red numbers denote the Chern number calculated in the $(k_y,\phi)$ or $(k_x,\phi)$ space using the states of the corresponding bands for each $k_x$ or $k_y$. (d) The spatial profile of the zero-energy corner modes. Here, $t=1$, $\lambda_x=\lambda_y=0.8$, and $\alpha_x=\alpha_y=1/4$.}
\label{fig2}
\end{figure}

If we take $\phi$ as the momentum $k_z$ along the $z$ direction, the 2D AAH model can be transformed into a 3D model
(the corresponding tight-binding Hamiltonian can be found in the Supplemental Material). This 3D model describes a 3D topological semimetal at half
filling, where gapless points exist on both $x$-normal and $y$-normal surfaces [see Figs.~\ref{fig1}(b) and \ref{fig1}(c)].
The zero-energy corner modes then contribute to the zero-energy hinge modes in the semimetal.
In addition, the system exhibits a surface Chern
band localized on the $x$-normal surfaces, leading to the chiral hinge modes.

The nonzero-energy corner modes (or chiral hinge modes in 3D) are in fact rooted in the topological properties of bulk energy bands.
To illustrate this, we calculate the Chern number of the Wannier bands, which are obtained based on
the lowest two degenerate occupied bands~\cite{SM}, and find that the Chern number is equal to one, which is consistent with
the existence of one chiral mode at each corner. Since this topological invariant is calculated using the wave functions under periodic boundary conditions, it shows that the nonzero-energy topological modes arise from the topological properties of
bulk states.

Another important observation of this model is that such nonzero energy corner modes can exist in the continuous 
bulk spectrum as indicated by the white circle in Fig.~\ref{fig1}(b).
These corner states lying in the continuum of the extended bulk states form the BICs. Such BICs also survive in the incommensurate lattice where the chiral symmetry is broken, as shown in Fig.~\ref{fig1}(j). Therefore, our 2D AAH model also 
provides a platform for studying the BICs in higher-order topological systems.

\emph{Case $\alpha_x=\alpha_y=1/4$}.--- We now investigate the case where the modulations have the same period in both directions,
e.g., $\alpha_x=\alpha_y=1/4$. In Fig.~\ref{fig2}(a), we display the energy spectrum for a lattice under open boundary conditions along both directions. Again, the zero-energy modes show up for $\phi \in (-\frac{3\pi}{4}, -\frac{\pi}{4})$ and $(\frac{\pi}{4}, \frac{3\pi}{4})$, which coincide with the regime for the emergence of zero-energy edge modes in the 1D AAH model. The zero-energy modes are fourfold degenerate corresponding to the corner states shown in Fig.~\ref{fig2}(d). Similar to the preceding case, the zero-energy corner modes are characterized by
the quantized quadrupole moment, as shown in Fig.~\ref{fig2}(b). In this case, the quadrupole moment arises from the bulk energy gap
closure [see Fig.~\ref{fig2}(c)], instead of the edge gap closure. These topological quadrupole insulators are also
type-I since $p_x^{\textrm{edge}}=p_y^{\textrm{edge}}=e/2$.

In addition to the zero-energy corner modes, we also find nonzero energy chiral modes inside the gaps of the energy spectrum.
One group of chiral modes in the bulk gaps are boundary bands with states localized at the 1D edges. They are characterized by the Chern number computed
in the $(k_y,\phi)$ or $(k_x,\phi)$ space using Eq.(\ref{Chern}) for each $k_x$ or $k_y$,
as denoted by the red numbers in Fig.~\ref{fig2}(c). The other chiral modes, marked by the red lines in Fig.~\ref{fig2}(a), connect different boundary bands and are localized at the corners.
Compared to the preceding case, the Chern number of the boundary states is not well defined
since these bands are chiral and not well separated.
Similarly, we can transform the 2D model into a 3D model with gapless points in the bulk [see Fig.~\ref{fig2}(c)], where
the zero-energy and nonzero energy corner states become the zero-energy and nonzero-energy chiral
modes localized at the hinges along $z$.

\emph{Experimental realization}.--- Here, we propose an experimental scheme to realize the 2D AAH model with electric circuits. 
In fact, the quadrupole topological insulators with zero-energy corner modes have been experimentally observed in electric circuits~\cite{Imhof2018Nat}. We apply a similar method to construct an electric network to simulate the 2D AAH model
using the Laplacian of the circuit
(see the details in the Supplemental Material). The quadrupole moment can be measured by probing the single-point impedances of the circuit~\cite{Xu2019arxiv} and
the existence of the corner modes can be detected by measuring the resonance of two-point impedances near the corners~\cite{Imhof2018Nat,Zeng2020PRB,Xu2019arxiv}.
In addition, we can use the electric network to simulate the dynamics of the
Schr\"odinger equation. In this case, the dynamics is governed by a circuit Hamiltonian (usually different from
the Laplacian) which can also support the zero-energy corner modes and nonzero-energy chiral modes.
Remarkably, we find the presence of chiral corner modes in the continuous bulk spectrum, indicating 
the existence of BICs in the electric circuit. Besides the electric circuit, our model can also be realized in other systems, such as
solid-state materials, cold atoms, and photonic and phononic crystals. 

In summary, we have constructed a generalized 2D AAH model that supports the coexistence of zero-energy corner modes characterized by the quantized quadrupole moment
and nonzero-energy corner modes characterized by the Chern number of the boundary bands and the Chern number of
the Wannier bands. This model actually describes a 3D higher-order topological semimetal with the coexistence
of zero-energy hinge modes and nonzero-energy chiral modes contributed by a Chern band on the surfaces. In addition, we also find that nonzero-energy chiral modes can form the bound states in the continuum.
Finally, we propose a practical scheme to realize the 2D AAH model in electric circuits. Our model provides another platform for realizing and detecting the exotic properties of higher-order topological insulators and semimetals.

\begin{acknowledgments}
This work is
supported by the start-up fund from Tsinghua University,
the National Thousand-Young-Talents Program and the National Natural Science Foundation
of China (11974201).

\emph{Note added}. Recently, we became aware of a related work on higher-order topological insulators~\cite{Zilberberg2019arXiv}.
\end{acknowledgments}

\begin{widetext}
\section{Supplemental Material}
\setcounter{equation}{0} \setcounter{figure}{0} \setcounter{table}{0} %
\renewcommand{\theequation}{S\arabic{equation}}
\renewcommand{\thefigure}{S\arabic{figure}}
\renewcommand{\bibnumfmt}[1]{[S#1]}

In the supplementary material, we will show how the boundary bands are extracted,
provide the 3D tight-bind Hamiltonian in real space obtained by transforming the 2D AAH model,
present the method to evaluate the Chern number of Wannier bands and introduce the detailed experimental scheme for realizing the 2D AAH model in electric circuits.

\section{S1. The boundary Chern band}
To show the existence of a Chern band at the edge, we plot the energy spectrum of the 2D AAH model
in a cylinder geometry with open and periodic boundaries along $x$ and $y$, respectively, in Fig.~\ref{figS1}(a).
To distinguish the bulk and edge modes, we calculate the inverse participation ratio (IPR) for each eigenstates $|\Psi_n \rangle$ with the corresponding eigenenergies $E_n$, which is defined as
\begin{equation}
\textrm{IPR} = \sum_{j} \frac{|\Psi_{n,j}|^4}{(\langle \Psi_n | \Psi_n \rangle)^2},
\end{equation}
with $\Psi_{n,j}$ being the $j$th component of the eigenstate vector. The IPR values of the edge modes are much larger than the bulk states, as shown in Fig.~\ref{figS1}(a). We therefore can extract the boundary bands based on the IPR values.
Specifically, we take $\phi=0$ and pick out the eigenstates with IPR values larger than a threshold value, say, for example, 0.025, and get the indices of these states. The boundary bands are composed of the states with the same indices at each $\phi$ value [Fig.~\ref{figS1}(b)]. After that, we calculate the Chern number of these boundary bands as presented in the main text.

\section{S2. The tight-binding model in 3D}
If we regard the parameter $\phi$ as the momentum $k_z$ along $z$, we obtain a Hamiltonian describing a 3D
system. By performing the Fourier transformation, we obtain the following tight-binding Hamiltonian in real space
\begin{eqnarray}\label{H_3d}
  H_{3D} &=&\sum_{j_x,j_y,j_z}\left[t_x (-1)^{j_y} \hat{c}_{j_x,j_y,j_z}^\dagger \hat{c}_{j_x+1,j_y,j_z} + t_y \hat{c}_{j_x,j_y+1,j_z}^\dagger \hat{c}_{j_x,j_y,j_z} \right.\nonumber \\
  && +\frac{t_x\lambda_x}{2}(-1)^{j_y} e^{i 2\pi\alpha_x j_x} (\hat{c}_{j_x,j_y,j_z}^\dagger \hat{c}_{j_x+1,j_y,j_z+1} + \hat{c}_{j_x+1,j_y,j_z}^\dagger \hat{c}_{j_x,j_y,j_z+1}) \nonumber\\
  && \left. +\frac{t_y\lambda_y}{2} e^{i 2\pi\alpha_y j_y} (\hat{c}_{j_x,j_y+1,j_z}^\dagger \hat{c}_{j_x,j_y,j_z+1} + \hat{c}_{j_x,j_y,j_z}^\dagger \hat{c}_{j_x,j_y+1,j_z+1}) + H.c.\right].
\end{eqnarray}

\begin{figure}[t]
  \includegraphics[width=5.0in]{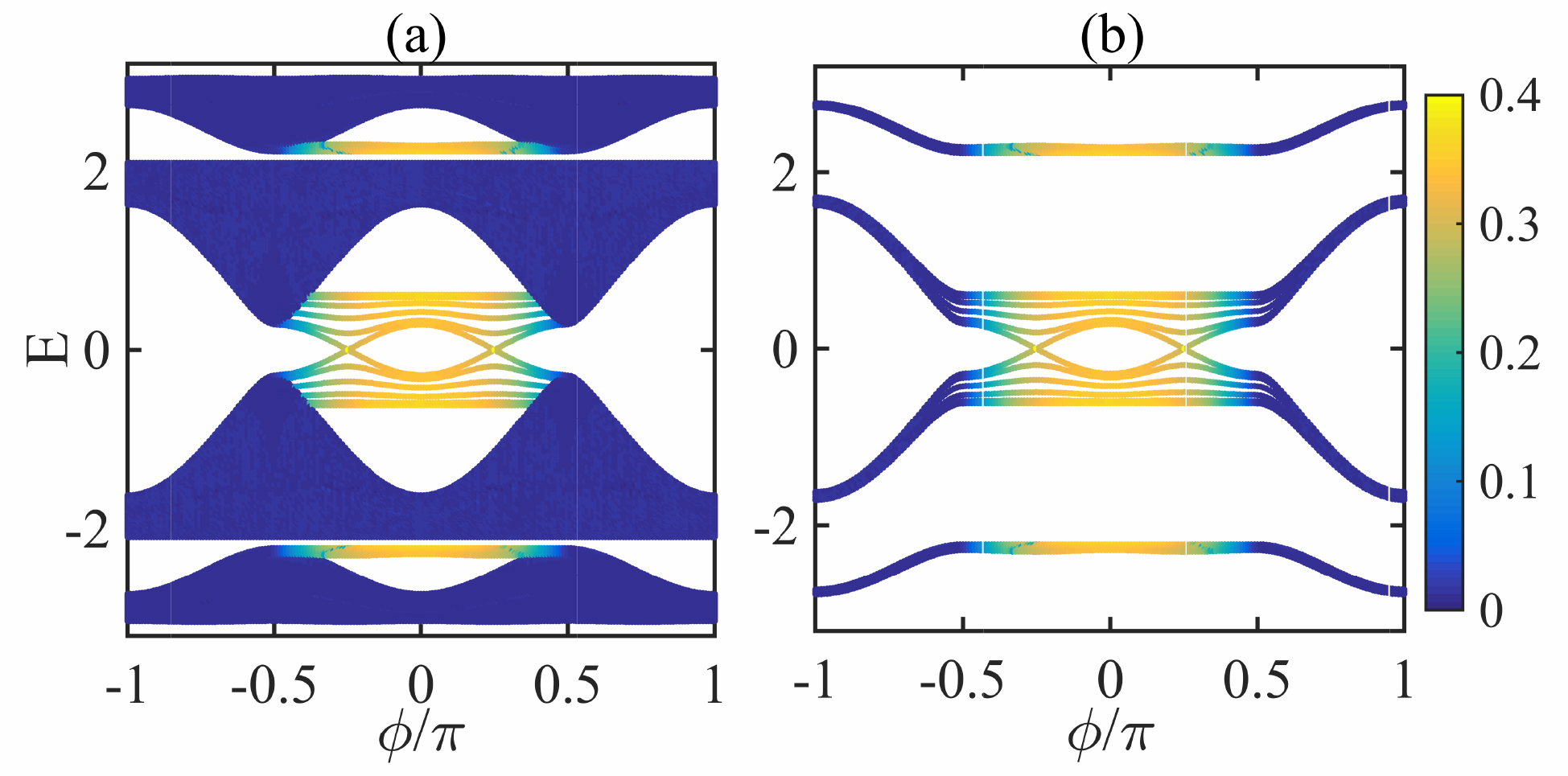}
  \caption{(Color online) (a) The energy spectrum of the 2D AAH model with open boundaries along $x$ and periodic boundaries
  along $y$. (b) The boundary bands extracted from the energy spectrum in (a). The color bar indicates the IPR value of the corresponding eigenstates. Here $t=1$, $\lambda_x=\lambda_y=0.8$, $\alpha_x=1/2$ and $\alpha_y=1/4$. The lattice size is $40 \times 40$.}
\label{figS1}
\end{figure}

\section{S3. The Chern number of the Wannier bands}
In the case of $(\alpha_x,\alpha_y)=(1/2,1/4)$, we first calculate the Wilson loop matrix along the $x$ direction for the
lowest doubly degenerate bands (separated from the other bands)~\cite{Taylor2017Science,Taylor2017PRB},
\begin{equation}
\mathcal{W}_{x,\bm{k}}=F_{x,\bm{k}+(N_x-1)\bm{\delta k_x}}\cdots F_{x,\bm{k}+\bm{\delta k_x}} F_{x,\bm{k}},
\label{WS1}
\end{equation}
where $[F_{x,\bm{k}}]^{mn}=\langle u_{\bm{k}+\bm{\delta k_x}}^m|u_{\bm{k}}^n\rangle$ and $\bm{\delta k_x}=(2\pi/N_x,0)$ with $|u_{\bm{k}}^n\rangle$ being the occupied eigenstate.
The Wannier Hamiltonian $H_{\mathcal{W}_{x,\bm{k}}}$ is defined as
\begin{equation}
\mathcal{W}_{x,\bm{k}}=e^{i H_{\mathcal{W}_{x,\bm{k}}}},
\end{equation}
which is related to the boundary bands with the edge perpendicular to the $x$ direction~\cite{Klich2011PRL}.

Using the spectral decomposition, the Wilson loop matrix $\mathcal{W}_{x,(\bm{k},\phi)}$ can be written as
\begin{equation}\label{Wannier}
\mathcal{W}_{x,(\bm{k},\phi)}=\sum_{j=\pm} e^{2\pi i \nu_x^j(k_y,\phi)} |\nu_{x,(\bm{k},\phi)}^j\rangle \langle\nu_{x,(\bm{k},\phi)}^j|,
\end{equation}
where $|\nu_{x,(\bm{k},\phi)}^j\rangle$ is the eigenvector of $\mathcal{W}_{x,(\bm{k},\phi)}$ with
the corresponding eigenvalue of $e^{2\pi i \nu_x^j(k_y,\phi)}$. We refer to the eigenvalues of $\nu_x^j(k_y,\phi)$
as the Wannier bands. Since $e^{2\pi i \nu_x^j(k_y,\phi)}$ repeats over $1$ for $\nu_x^j(k_y,\phi)$,
we restrict them to $[0,1)$. As shown in Fig.~\ref{figS2}, there appear two Wannier bands that are gapped around $\nu_x=0.5$,
leading to two Wannier sectors denoted by $\nu_x^{-}$ for the band below the Wannier gap and $\nu_x^{+}$ for the band
above the Wannier gap.

Analogous to the Wannier-sector polarization~\cite{Taylor2017Science,Taylor2017PRB},
we define the Wannier-sector Chern number using the Wannier basis~\cite{Bernevig2018SA}
\begin{equation}
C^{\nu_x^{\pm}}_{k_y,\phi}=\frac{1}{2\pi}\frac{1}{N_x}\sum_{k_x}\int_{0}^{2\pi}d \phi\int_{0}^{2\pi}d k_y \tilde{\Omega}^{\pm}_{k_x,(k_y,\phi)},
\end{equation}
where $\tilde{\Omega}^{\pm}_{k_x,(k_y,\phi)}=i(\langle\partial_{k_y} w^{\pm}_{x,(\bm{k},\phi)}|\partial_{\phi} w^{\pm}_{x,(\bm{k},\phi)}\rangle - k_y\leftrightarrow \phi )$ and
\begin{equation}
|w^{\pm}_{x,(\bm{k},\phi)}\rangle=\sum_{n=1,2}|u_{\bm{k}}^n\rangle [\nu_{x,(\bm{k},\phi)}^j]^n.
\end{equation}
In the thermodynamic limit, the Chern number of the Wannier band is
\begin{equation}
C^{\nu_x^{\pm}}_{k_y,\phi}=\frac{1}{4\pi^2}\int_{0}^{2\pi}d\phi \int_{BZ}d^2\bm{k} \tilde{\Omega}^{\pm}_{k_x,(k_y,\phi)}.
\end{equation}

\begin{figure}[t]
  \includegraphics[width=3.4in]{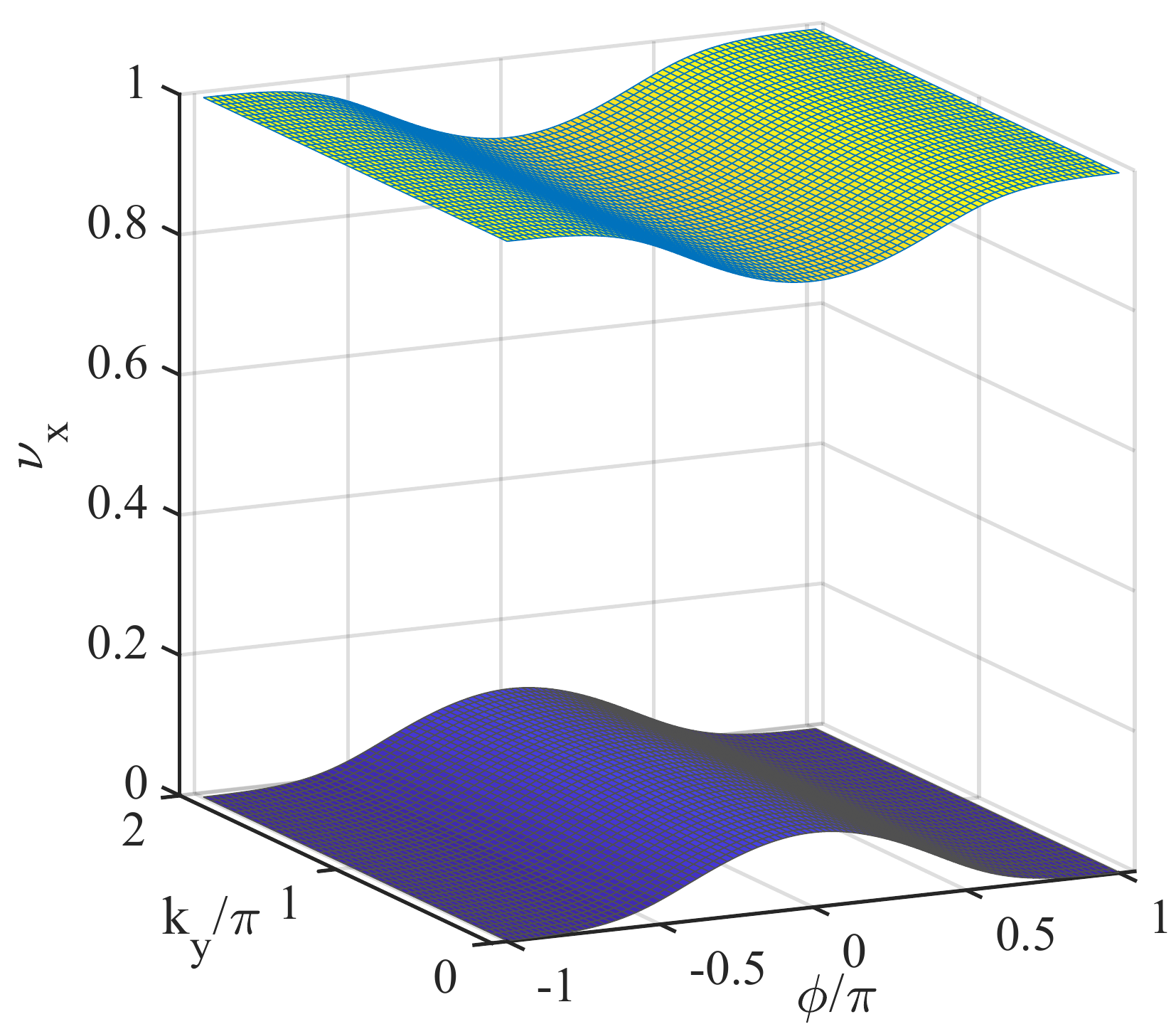}
  \caption{(Color online) The Wannier bands for the system with $\alpha_x=1/2$ and $\alpha_y=1/4$ obtained by diagonalizing the Wilson loop matrix [see the formula (\ref{WS1})]. Here $t=1$ and $\lambda_x=\lambda_y=0.8$.}
\label{figS2}
\end{figure}

For the lowest two occupied bands, we find that the Chern numbers of Wannier sectors are $C^{\nu_x^{+}}_{k_y,\phi}=C^{\nu_x^{-}}_{k_y,\phi}=1$,
which characterize the chiral corner modes. This topological invariant is equivalent to the Chern number of boundary bands defined in the main text.

\section{S4. Experimental realization in electric circuit}
In this section, we present the experimental scheme in detail for realizing the 2D AAH model in electric circuit.
Fig.~\ref{figS3}(a) illustrates an electric network composed of capacitors and inductors. This network can be used to
simulate our 2D Hamiltonian with lattice sites described by the circuit nodes indicated by red indices.
If we denote the input currents and voltages at each node as a $N$-component column vector $\bm{I}$ and $\bm{V}$, respectively,
it is straightforward to obtain a relation expressed as $\boldsymbol{I}=J \boldsymbol{V}$, according to the Kirchhoff's law.
Here $J$ is the Laplacian of the circuit. Our Hamiltonian can be simulated by the Laplacian
through $J=iH$. Specifically, the connection between the neighboring nodes through inductors or capacitors can imitate the hopping between neighboring lattice sites. If the hopping is negative (positive), the two nodes are connected by an inductor (a capacitor). For a specific node $(m,n)$, we set the impedance of the device as $1/Z_{xm}= i t [1+\lambda_x \cos (2\pi \alpha_x m + \phi_x)]$ and $1/Z_{yn}=it[1+\lambda_y \cos (2\pi \alpha_y n + \phi_y)]$ (here $i$ refers to the imaginary unit).
In addition, we need to eliminate the diagonal terms by grounding every node through a capacitor (inductor) with appropriate impedance $1/Z^\prime_{m,n}=-(1/Z_{xm}+1/Z_{x,m-1}+1/Z_{yn}+1/Z_{y,n-1})$. 

To study the bound states in the continuum (BICs) which are stable against scattering to the bulk states,
we need to investigate the dynamics in electric circuits. In fact, the dynamics of
a wave packet in a Chern insulator~\cite{Thomale2019PRL} and the dynamics of a hinge soliton in a 3D higher-order
topological insulator~\cite{Tao2020arXiv} have been studied in electric circuits.
The time evolution of the circuit is governed by~\cite{Thomale2019PRL,Tao2020arXiv}
\begin{equation}
\frac{d}{dt}\bm{I}(t)=C\frac{d^2}{dt^2}\bm{V}(t)+\Sigma\frac{d}{dt}\bm{V}(t)+L\bm{V}(t),
\end{equation}
where $C$, $\Sigma$, and $L$ are real-valued matrices for capacitance, conductance, and inductance, respectively.
The circuit Laplacian for alternating currents with frequency $\omega$ is given by
\begin{equation}
J(\omega)=i\omega C + \Sigma + \frac{1}{i\omega} L.
\end{equation}

If there are no input currents ($\bm{I}(t)=0$), the equation of motion of the circuit
can be reduced to a first-order differential equation which is similar to the Schr\"{o}dinger equation,
\begin{equation}
-i\frac{d}{dt}\psi(t)=H_{c}\psi(t),
\end{equation}
where $\psi(t)=(\frac{d}{dt}{{\bm{V}}}(t),\bm{V}(t))^T$, and the circuit Hamiltonian is
\begin{equation}
H_{c}=i\begin{pmatrix}
C^{-1}\Sigma & C^{-1}L \\
-\mathds{1} & 0 \end{pmatrix}.
\end{equation}
The time evolution of an eigenstate of $H_{c}$ is given by $\psi_{n}(t)=e^{i\omega_n t}\phi_n$,
where $\phi_n$ is an eigenvector of $H_{c}$ with the eigenfrequency $\omega_n$ ($n=1,\cdots,2N$).
It can be shown that the eigenfrequencies of $H_{c}$ occur in real-valued pairs $(\omega_n,-\omega_n)$
if the Laplacian matrix $J(\omega)$ satisfies $J(\omega)=-J(\omega)^{\dagger}$ for any real frequencies $\omega$~\cite{Thomale2019PRL}.
These eigenfrequencies $\{\omega_n\}$ of the circuit Hamiltonian $H_c$ are also associated with the admittance eigenvalues $j(\omega)$
of the Laplacian $J(\omega)$ through $j(\omega_n)=0$.
Due to this intimate connection, the circuit Hamiltonian $H_c$ owns similar band structures to those of the circuit
Laplacian $J$~\cite{Thomale2019PRL}. For instance, if there exist four zero-energy corner modes in 
the eigenvalues of $J(\omega_0)=i\omega_0 H(\omega_0)$, i.e., $j_n(\omega_0)=0$ with $n=1,2,3,4$, associated with
the corresponding eigenvectors $V_n$, $H_c$ also has four-fold degenerate frequency modes $(\begin{array}{cc}
                                                                               i\omega_0V_n & V_n 
                                                                             \end{array})^T
$ corresponding to the frequency $\omega_0$.

\begin{figure}[t]
  \includegraphics[width=5.0in]{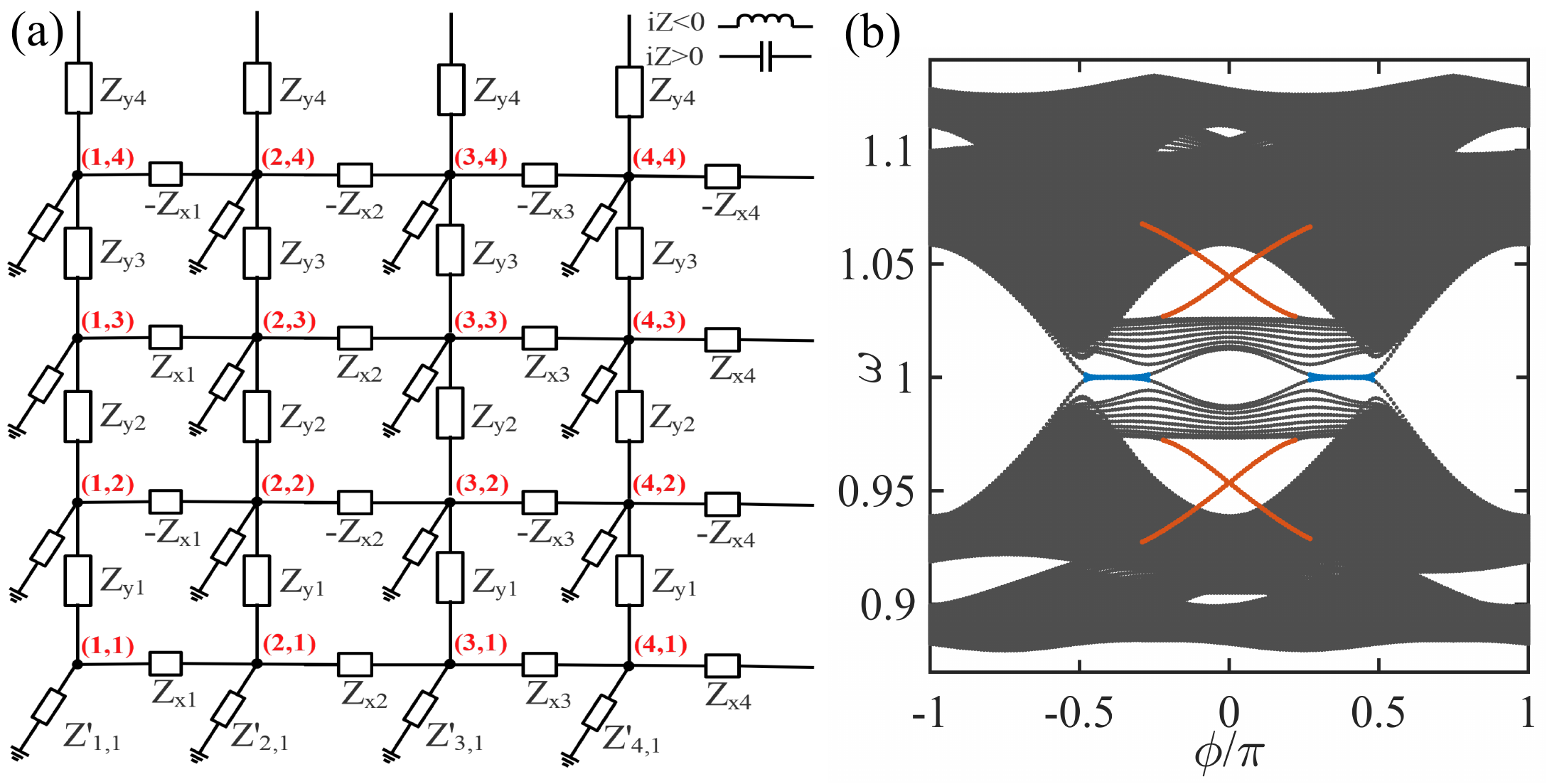}
  \caption{(Color online) (a) The electric network for realizing the 2D AAH model. The red numbers in brackets label
  the indices of nodes corresponding to the lattice sites. The hopping between the neighboring lattice sites are simulated by the impedance of the electric device connecting the nodes. If the hopping amplitude is positive (negative), then $Z$ is implemented by a capacitor (an inductor). Every node is grounded through a capacitor or an inductor to eliminate the diagonal terms. (b) The frequency spectrum $\omega$ of the circuit with respect to $\phi$ for the 2D AAH model with $\alpha_x=1/2$ and $\alpha_y=1/4$ in a
  geometry with open boundaries in all directions. The unit of $\omega$ is $\omega_0$.}
\label{figS3}
\end{figure}

In Fig.~\ref{figS3}(b), we plot the eigenfrequencies $\omega$ of $H_c$ as a function of $\phi$ for the case with $\alpha_x=1/2$ and $\alpha_y=1/4$ in a geometry
with open boundaries along both $x$ and $y$ directions.
As expected, the frequency bands $\omega(\phi)$ exhibit four-fold degenerate corner modes with the eigenfrequency 
equal to the resonance frequency $\omega_0$, as shown by the blue lines in the figure. We also find chiral modes (red lines) localized at the corners as well, which is similar to the energy spectrum of the Hamiltonian $H$. 
It is interesting to see that parts of the chiral corner modes lie in the continuous bulk spectrum, forming 
BICs.
\end{widetext}

\end{document}